# LONG-TERM TRENDS IN THE SOLAR WIND PROTON MEASUREMENTS

Heather A. Elliott[1], David J. McComas[1, 2], and Craig E. DeForest[3]

[1] Southwest Research Institute, San Antonio, Texas, USA

[2] Princeton, Princeton University, Princeton, New Jersey, USA

[3] Southwest Research Institute, Boulder, Colorado, USA

## ABSTRACT

We examine the long-term time evolution (1965-2015) of the relationships between solar wind proton temperature (Tp) and speed ($V_p$) and between the proton density ($n_p$) and speed using OMNI solar wind observations taken near Earth. We find a long-term decrease in the proton temperature-speed ($T_p$-$V_p$) slope that lasted from 1972 to 2010, but has been trending upward since 2010. Since the solar wind proton density-speed ($n_p$-$V_p$) relationship is not linear like the $T_p$-$V_p$ relationship, we perform power law fits for $n_p$-$V_p$. The exponent (steepness in the $n_p$-$V_p$ relationship) is correlated with the solar cycle. This exponent has a stronger correlation with current sheet tilt angle than with sunspot number because the sunspot number maxima vary considerably from cycle to cycle and the tilt angle maxima do not. To understand this finding, we examined the average $n_p$ for different speed ranges, and found that for the slow wind $n_p$ is highly correlated with the sunspot number with a lag of ~4 years. The fast wind $n_p$ variation was less, but in phase with the cycle. This phase difference may contribute to the $n_p$-$V_p$ exponent correlation with the solar cycle. These long-term trends are important since empirical formulas based on fits to $T_p$ and $V_p$ data are commonly used to identify ICMEs, but these formulas do not include any time dependence. Changes in the solar wind density over a solar cycle will create corresponding changes in the near Earth space environment and the overall extent of the heliosphere.

*Key words:* interplanetary medium – solar wind – Sun: heliosphere – Sun: solar-terrestrial relations

## 1. INTRODUCTION

Some of the earliest in situ observations of the solar wind found a strong correlation between the solar wind proton temperature ($T_p$) and speed ($V_p$) measurements (Neugebauer & Snyder 1966). The proton temperature-speed ($T_p$-$V_p$) relationship has shown to be a reliable method for identifying Interplanetary Coronal Mass Ejections (ICMEs) (Lopez & Freeman 1986; Richardson & Cane 1995; Cane & Richardson 2003; Richardson & Cane 2010; Elliott et al. 2005; Neugebauer et al. 2003). The strong correlation between the solar wind proton temperature and speed breaks down for wind in transient ICMEs, and a majority of ICMEs are much colder than what would be expected for non-transient wind moving at the same speed as the ICME (Matthaeus et al. 2006; Gosling et al. 1972; 1973). When a large collection of in situ solar wind proton temperature measurements are plotted versus the corresponding speed measurements, most of the points lie along a common curve with a steep slope where the temperature steadily increases as the speed increases. In contrast, the ICMEs create a horizontal line at a low temperature (Matthaeus et al. 2006). Given that the $T_p$-$V_p$ relationship is often used to identify ICMEs, any temporal variations in the relationship will impact the utility of any given empirical temperature-speed formula. The relationship between the solar wind proton density ($n_p$) and speed has not been examined in the same detail as the $T_p$-$V_p$ relationship because $n_p$ and $V_p$ are not as well correlated with one another, and the density-speed ($n_p$-$V_p$) relationship has not been found to be as useful for identifying ICMEs. The solar wind proton density is highly variable and often displayed on a logarithmic scale. Solar rotation averages of the proton density can vary by an order of magnitude (McComas et al. 2008), and hourly averages can vary by nearly 2 orders of magnitude (McComas et al. 2002). Similarly the average proton density for a Interplanetary Coronal Mass Ejection can vary by more than one order of magnitude (Siscoe et al. 2006). One type of study where the density and speed are both analyzed are in superposed epoch studies of Corotating Interaction Regions (CIRs) (e.g. McPherron and Weygand (2006); Borovsky and Denton (2010)), but typically no empirical expression is provided for the $n_p$-$V_p$ relationship as is common in the $T_p$-$V_p$ relationship studies mentioned above.

Ecliptic solar wind observations of the solar wind temperature and speed at 1 AU have been fit with linear, quadratics, and polynomial expressions up to the second order (Lopez & Freeman 1986; Richardson & Cane 1995; Neugebauer et al. 2003; Burlaga & Ogilvie 1973; 1970; Hundhausen et al. 1970). Elliott et al. (2012) examined the two most common functional forms linear and quadratic, and found the linear fits work better. Several early studies found a break in the $T_p$-$V_p$ plot where the slow and fast wind followed different curves (e.g. Neugebauer et al., (2003)). Elliott et al. (2012) found that in ecliptic 1 AU observations such a break was difficult to detected for most years because there were not many data points above 570 km $s^{-1}$. However, in 2003 they found a distinct break when a large low latitude extension to a polar coronal hole lasted about a year and produced plenty of non-transient wind measurements extending to ~800 km $s^{-1}$. They also found that when such a distinct break is present separate linear fits for the slow and fast wind characterized the $T_p$-$V_p$ distribution well.

This paper extends the analysis of the time evolution of solar wind proton temperature and speed relationship by Elliott et al., (2012), and performs a similar analysis for the solar wind proton density and speed relationship. We find that the long-term decrease in the $T_p$-$V_p$ slope that began in 1972 and ended in 2010, and since 2010 the slope has been increasing. The solar wind $n_p$-$V_p$ relationship is clearly not linear like the $T_p$-$V_p$ relationship; therefore, we perform power law fits for $n_p$-$V_p$. The power law exponent representing the steepness in the $n_p$-$V_p$ relationship is correlated with the solar cycle. Both the sunspot number and heliospheric current sheet tilt angle exhibit an ~11 year solar cycle, but the amplitude for each cycle varies more for the sunspot number than for the tilt angle. We find the $n_p$-$V_p$ power law exponent has a stronger correlation with the current sheet tilt than the sunspot number.

## 2. OBSERVATIONS

To examine the evolution of the $T_p$-$V_p$ and $n_p$-$V_p$ relationships over time scales spanning multiple solar cycles, we primarily use the hourly OMNI 2 data set available from the National Space Science Data Center (NSSDC), which begins on 1 January 1963. The temperature and density measurements begin in 1965, and our statistical analysis goes through the end of 2015. King and Papitashvili [2005] describe this collection of solar wind observations from many spacecraft in the solar wind near Earth, and detailed documentation is online (http://omniweb.gsfc.nasa.gov). To compensate for some differences in calibrations, the data were normalized by examining times when missions overlapped. In a prior study we independently validated these normalizations (Elliott et al. 2012). The other type of adjustment performed in creating the OMNI dataset is that measurements taken away from Earth have been time shifted to projected times at Earth. We also compare the trends found in the OMNI solar wind observations to those from the Advanced Composition Explorer (*ACE*), which began on January 28, 1998. The hourly level 2 *ACE* Solar Wind Electron, Proton, and Alpha Monitor (SWEPAM) electrostatic analyzer measurements were obtained from the ACE science center (McComas et al. 1998).

Since we are interested in the properties of the non-transient wind, we remove the Interplanetary Coronal Mass Ejections (ICMEs). We remove times in the Richardson and Cane ICME list (Cane & Richardson 2003; Richardson & Cane 2010; 2004), which is described further online as a level 3 data product at the *ACE* science center. To exclude compressions associated with ICMEs, and to allow for some uncertainty in the stop and start times, we remove all points that range from 15 hours prior to the start times through 6 hours after the stop times. The Richardson and Cane list begins in 1996 and does not cover the early OMNI observations, but does cover the *ACE* observations. Therefore, we also excluded times when either the proton plasma beta is low ($\beta_p < 0.1$), or the alpha ($\alpha$) to proton density ratio is high ($n_\alpha/n_p > 0.08$) along with any data within 15 hours prior and 6 hours after either of those criteria being satisfied for both OMNI and *ACE*. For the earliest measurements, we rely on the proton beta criteria to remove the ICMEs since the alpha data begins in 1971. In Elliott et al. (2012), we demonstrated that this technique is successful at removing the ICMEs since no significant anomalous low temperature population remain.

## 2. RESULTS

Our study extends the analysis in Elliott et al. (2012) of the temporal evolution of the temperature-speed distribution from 1965 through 2009. We add observations from 2010 through the end of 2015 and have updated to the latest version of the OMNI combined dataset since several revisions to the prior OMNI data had occurred in the interim. Each panel in Figure 1 shows a 2-dimensional binned color distribution plot of all non-ICME hourly OMNI temperature and speed measurements for each year from 2010 to 2015. We performed several different linear fits to the individual hourly samples. In Figure 1 we show the fit results to all the observations between 330 and 850 km $s^{-1}$ as was done in Elliott et al. (2012). We maintain this limit both for comparison purposes and because the rational for this lower limit is still valid. This low speed limit was originally set because some years do not have many points below 330 km $s^{-1}$, and it is more difficult to remove very slow ICMEs since they have properties similar to the slow wind. Any points above 850 km $s^{-1}$ are associated with fast ICMEs. However, later we show slope results from additional fits over other speed ranges. The majority of the observations in Figure 1 for each year have speeds less than 450 km $s^{-1}$ and would be classified as slow wind.

In Figure 2 we show that the updated $T_p$-$V_p$ slope time evolution. Our main finding is that the long decrease in the proton temperature-speed slope that began in 1972 lasted a little more than 3 solar cycles and ended in 2010. The top panel shows a black curve that corresponds to the fits for speeds between 330 and 850 km $s^{-1}$ for all non-ICME hourly OMNI observations. The blue curve is the same analysis done for the hourly *ACE* solar wind observation. In red are the fits to the OMNI observations for speeds ≥ 570 km $s^{-1}$ when the number of points with speeds ≥ 570 km $s^{-1}$was at least 170. Similarly, orange plus signs show the *ACE* observations for the fast wind with the same number points required. The middle panel shows the same kind of analysis and fits for additional speed ranges. In purple we show slope results for fits in the speed range from 330 to 450 km $s^{-1}$ with the same number of points requirement. This range encompasses all the slow wind observations where we have a significant number of points. The corresponding results for *ACE* are shown in pink. The slope results for fits to measurements with speeds ≤ 450 km $s^{-1}$ are in bright green for OMNI and in dark green for the *ACE*. Since many studies use all points below 450 km $s^{-1}$, we also fit all points below 450 km $s^{-1}$ although as mentioned earlier for some years there are not many points below 330 km s$^{-1}$. The top and middle panels can be compared to the sunspot number and current sheet tilt shown in the bottom pane of Figure 2. There was a long decrease in slopes for the fits to speeds between 330 and 850 km s$^{-1}$ shown in the top panel spanning from 1972 to 2010, which is a little more than 3 solar cycles (included as supplemental data files). Since 2010 the OMNI slopes have been rising and the rise is steeper in the *ACE* results. This time evolution of the slope in the 330-850 km s$^{-1}$ range is not easily characterized by a single curve. Therefore, we include these results as supplemental data files to allow the results to be incorporated into empirical models. The slow wind (middle panel) does not show the steep decrease from 1972 to 2010 found in the top panel.

We find no significant correlation with the solar cycle. In Figure 3 we plot the yearly OMNI slopes values versus the annual average sunspot number beginning with year 1973 since we confirmed the OMNI normalizations for observations beyond the beginning of 1973. We use the same color-coding as in Figure 2, and show the fit and the Pearson (Rp) and Sparkman (Rs) correlation coefficient results. We do not fit the fast wind points (red) since there were not enough of them. The fit results, and correlation coefficients results are shown in Table 1. The significance levels are shown for the Spearman correlation coefficient. For the Pearson correlations, we define Rcl to be the ratio of the Pearson correlation coefficient to $2/\sqrt{N}$. If the Pearson coefficient is $\leq 2/\sqrt{N}$, there is no significant correlation at the 95% confidence level when compared to a random distribution. When Rcl is ≥ 1.6 the correlation can be considered significant (Borovsky et al. 1998; Bendat & Piersol 2010; Elliott et al. 2001). The data in Figures 2 and 3 have been filtered to only show T-V slopes derived from fits where the Rcl value was > 1.6 and the magnitude of the Pearson correlation was > 0.3. From Table 1 we conclude and Figure 3 that for speeds ≥ 330 km s$^{-1}$ and speeds between 330 and 450 km s$^{-1}$, there is no significant correlation. However, there could be a weak correlation for the fits to speeds ≤ 450 km s$^{-1}$ shown in the middle panel. We examined the correlations shifting the slopes and sunspot numbers using lags +/- 5 years and did not find any stronger correlations than these.

The analysis on the density-speed ($n_p$-$V_p$) relationship is similar to what was done for the $T_p$-$V_p$ relationship, and we find a solar cycle dependence. A linear fit did not work well for the $n_p$-$V_p$ relationship. The density is a highly variable parameter and is often plotted on a log scale for this reason. Additionally the relationship between $n_p$ and $V_p$ is not as tight or as well defined as between $T_p$ and $V_p$ (Figure 4). However, the particle flux ($n_pV_p$) is known to be relatively constant for time scales greater than a solar rotation only slightly changing with time (McComas et al. 2013). We could not find a study of the $n_p$-$V_p$ relationship spanning a long time range. Therefore, we include an initial analysis in this study by performing power-law fits. Figure 4 shows annual plots from 2010 to 2015 of the hourly OMNI density and speed data placed into 2-D bins with logarithmic scaling for both axes. Figure 5 shows the time evolution of the exponent in the power law fit for the same speed ranges used earlier in the $T_p$-$V_p$ analysis shown in Figure 2. At low speeds we find the exponents show systematic changes over the course of a solar cycle. In Figures 6 and 7 we show the exponents plotted versus the yearly average sunspot number and current sheet tilt for the fit to data above 330 km s$^{-1}$ (black; top panel), below 450 km s$^{-1}$ (green; middle panel), between 330 and 450 km s$^{-1}$ (purple; bottom panel). Once again, we filtered the data to only show n-V exponents derived from fits where the Rcl value was > 1.6 and the magnitude of the Pearson correlation was > 0.3. Correspondingly, Tables 2 and 3 summarize the fit parameters, correlation coefficients, and significance levels for Figures 6 and 7. The highest correlation between the exponent and both the sunspot number (Rs=0.466) and tilt angle (Rp= 0.602) were for the speeds < 450 km s$^{-1}$. By comparing the results shown in Tables 2 and 3 it is clear that the correlation coefficients are lower and less significant for the exponent-sunspot relationship than for the exponent-tilt relationship particularly for the slow wind for speeds below 450 km $s^{-1}$. The number of points for the 330 km $s^{-1}$ and 450 km $s^{-1}$ speed range was significantly reduced by the requirements for Rcl value to be > 1.6 and the magnitude of the Pearson correlation to be > 0.3, and no clear between the exponents and either tilt or sunspot number can be discerned. This may indicate that the source of the wind the spacecraft encountered could depend on the current sheet tilt.

We further analyze the proton density to better understand the solar cycle dependence in the $n_p$-$V_p$ relationship. The solar cycle depenence for the $n_p$-$V_p$ relationship may not be surprising since the helium abundance level for the slow solar wind is known to vary over the solar cycle (Kasper et al. 2012). The helium abundance is defined as 100 times the ratio of alpha density ($n_\alpha$) to proton density. Therefore, we also calculated the average proton density for the speed ranges in our prior analysis and for slow wind speeds < 350 km $s^{-1}$, and the moderate wind from 450 to 570 km $s^{-1}$ (Figure 8) to facilitate comparisons with Kasper et al. (2012). To obtain an accurate fit you need more measurements over a wide range of speeds than you need to obtain an meaningful average. The slow wind shows the largest variation over a given solar cycle and depends on the strength of the solar cycle. Since the tilt angle maximum value does not vary much from cycle to cycle, we compare the average proton density to the sunspot number and not the current sheet tilt angle. The fast wind shows some variation with solar cycle, but to a lesser extent. The slow wind peaks occur in the mid to late declining phase. However the peaks for the fast wind occur near solar maximum or near the early declining phase. This phase difference for slow and fast wind may explain why the exponent for the $n_p$-$V_p$ relationship varies with the solar cycle. Since the slow wind helium abundance has a broad peak at solar maximum, the changes in the proton densities may be obscuring an even stronger relationship between the helium density and the solar cycle. The rise in proton density in the declining phase would cause the helium abundance to decrease more sharply if the helium level was constant. However, the helium abundance peak is broad. In the decent from the maximum abundance value in 2000 to the minimum in 2009, there is a bump in 2005 (Kasper et al. 2012), which happens to coincide to when the slow wind proton density drops significantly (Figure 8). The slow wind proton density seems to have a stronger relationship with the sunspot number and is not in phase with the solar cycle; therefore, we examine the density-sunspot relationship both without any time shifting, and with time shifting the sunspot number. We find that for the slow wind there is no significant correlation without time shifting (Figure 9 and Table 4), but a strong significant correlation (Rs=0.681 for $V \leq 350$ km $s^{-1}$) is found when the sunspot number is shifted forwards in time by 4 years (Figure 10 and Table 5).

After finding this result, we searched again for relationships with a similar phase difference and found the slow wind proton number flux ($n_pV_p$) had been found to have a correlation with the sunspot number ~4 years out of phase by Schwadron et al., (2011). However, based on those results it was not clear that the main contributor to the correlation between the proton flux ($n_pV_p$) and the sunspot number was the proton density and not the speed. The range of speeds of those flux calculations was narrow enough such that the main contributor to the relationship between the flux and sunspot was the density and not the speed. We also performed the same analysis for the proton number flux and dynamic pressure and find similar correlations as shown in Tables 6 and 7 for comparison purposes.

It seems reasonable that a solar cycle dependence would be either related to the solar sources of the wind, or the amount of energy available to accelerate and heat the wind. Theoretical work has shown that if energy is added at a low height the amount of mass flux in the solar wind is enhanced, but if added at a higher height it increases the solar wind speed (Leer & Holzer 1980). The solar cycle dependence could reflect a solar cycle dependence on the altitude at which the energy is added. The source of the slow solar wind is not well known [e. g. Kepko et al., (2016) and Crooker et al., (2012)]. The dominant source of the solar wind may change over the course of the cycle or either the source or sources may depend on the solar cycle. Xu and Borovsky (2015) found that below 400 km $s^{-1}$ the wind was predominately associated with either the streamer belts or sector reversal regions with the sector reversals being most important below 300 km $s^{-1}$. Their work shows the average density for sector reversal regions (12 $cm^{-3}$) is higher than their general streamer belt origin (6.3 $cm^{-3}$) category. They also show that the ratio of the sector reversal occurrence to the combined sector reversal and general streamer belt occurrence had a solar cycle dependence shifted by about 4 or 5 years. Given the Xu and Borovsky (2015) findings, multiple sources of the slow wind may explain the solar cycle dependence in the low speed wind proton density we observe.

## 4. Summary and Conclusions

The most significant findings from our statistical analysis of the non-transient ecliptic background solar wind are as follows.

1) The proton temperature-speed slope is now increasing after it had decreased for more than 3 solar cycles from 1972 to 2010.
2) The power law exponent for the proton density-speed relationship has a distinct solar cycle dependence particularly for slow solar wind speeds, and this dependence is more apparent when the current sheet tilt rather than the sunspot number is used as a measure of the solar cycle.

3) The average solar wind proton density for the slow solar wind is highly correlated with the sunspot number, but lags the peak of the solar cycle by about 4 years.

The empirical formulas based on fits to the temperature-speed observations are used to identify Interplanetary Coronal Mass Ejections (ICMEs). The formulas currently used do not include the time dependence that our study reveals. The helium abundance ratio ($100\times(n_p/n_\alpha)$) has a well-established solar cycle dependence in the slow solar wind. However, the solar cycle dependence in the slow wind proton density which is out of phase by ~4 years had not been explicitly shown. Prior studies focused on the proton flux and the helium abundance ratio. The proton density variations are the main reason the slow wind proton flux has a solar cycle dependence. Changes in the proton density with solar cycle also broaden the slow solar wind alpha abundance ratio enhancement that occurs at solar maximum; therefore, the slow wind helium density probably has a stronger solar cycle dependence than the helium abundance.

The proton density for the fast wind also has some solar cycle dependence in phase with the cycle. Therefore, although the solar wind flux ($n_pV_p$) is often thought to be relatively constant on solar rotation time scales (e.g. McComas et al., (2013)), the steepness in the $n_p$-$V_p$ relationship has a solar cycle dependence particularly for the slow solar wind. These relationships between the speed and both the temperature and density in the non-transient background solar wind are important because they can be used to estimate the density and temperature from either solar wind speed observations or predictions. Currently, the solar wind speed is the easiest solar wind parameter to predict based on solar observations and based on prior solar wind observations (Elliott et al. 2012). Therefore, defining clear relationships between the speed and both the density and temperature enables additional empirical forecasting and provides relationships, which the physical models must be able to reproduce. Changes in the solar wind density on the scale of the solar cycle will create corresponding changes in the near Earth space environment and overall extent of the heliosphere.

ACKNOWLEDGMENTS

We thank Natalia Papitashvilli and Joseph King for their work creating the OMNI collection of solar wind measurements. The work of Heather A. Elliott and David J. McComas is supported by the NASA SWEPAM MO&DA in Support of *ACE* grant NNX13AQ01G.

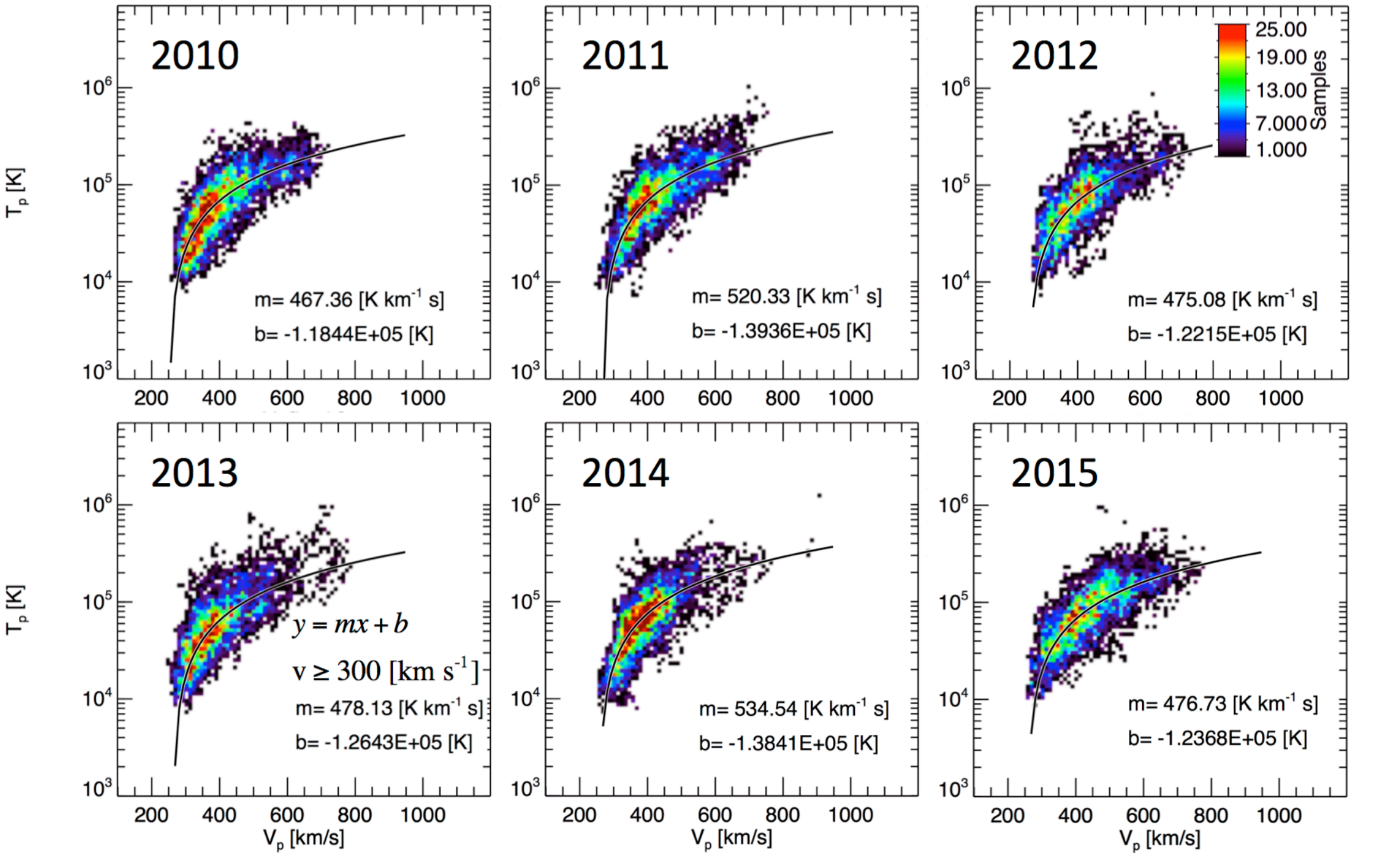

**Figure 1.** Temperature-speed distribution plots spanning from 2010-2015. After removing the ICMEs, the remaining hourly OMNI $T_p$-$V_p$ observations are placed in 2-D bins, and color coded by the number of points per bin. The fits are performed to all the hourly samples with the ICMEs removed, but without any binning. The black curves show linear fits over most of the energy range, and the fitting procedure is the same as those in *Elliott et al.,*(2012) where the measurements with speeds from 330 to 850 km s$^{-1}$ were fit.

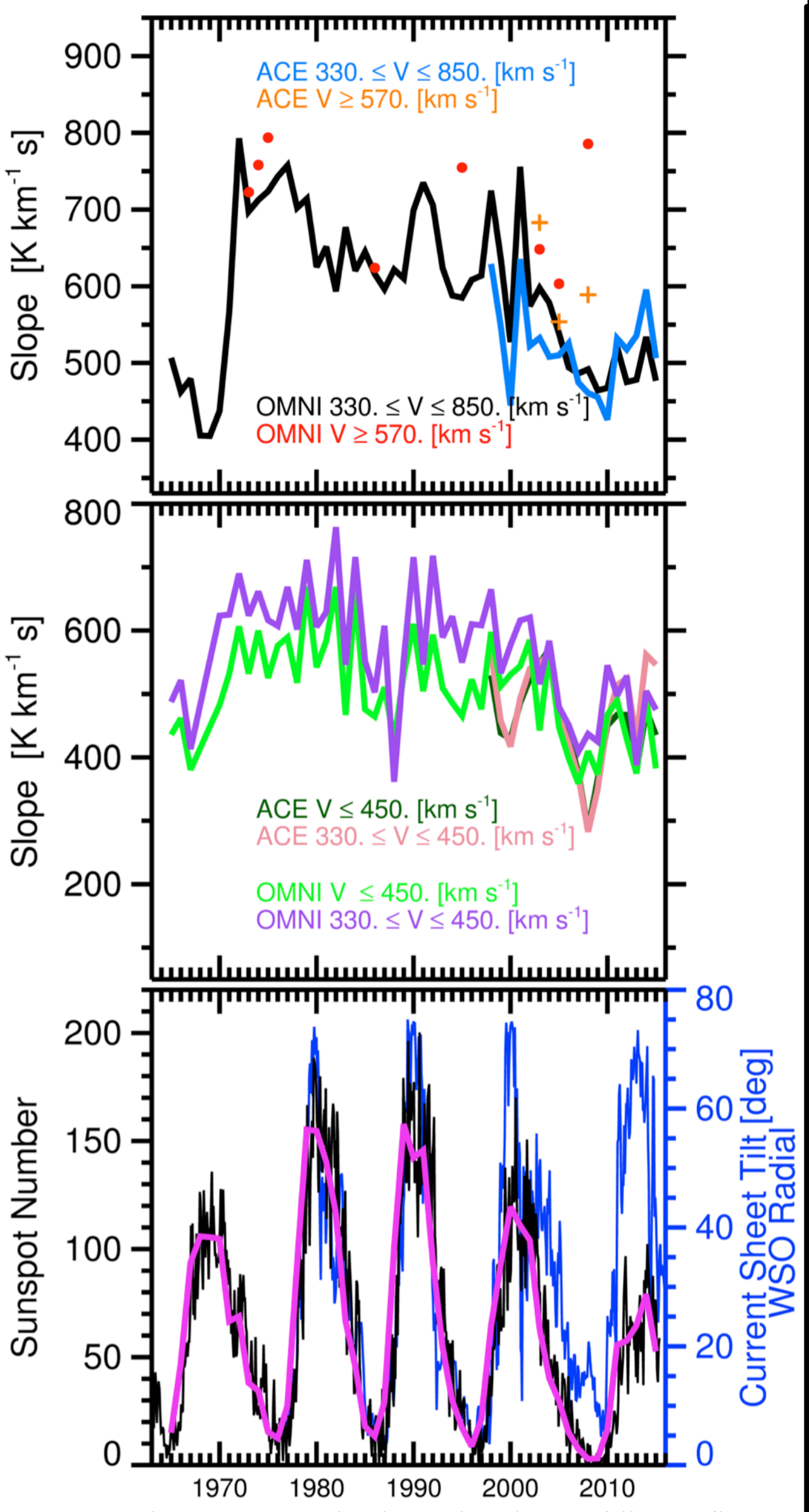

**Figure 2.** The top 2 panels show the slope of linear fits to hourly $T_p$ and $V_p$ observations for each year after having removed the ICMEs. In the top panel the slopes for OMNI data for speeds between 330 and 850 km $s^{-1}$ in black, for speeds above 570 km $s^{-1}$ in red. Similarly, the *ACE* slopes for speeds from 330 to 850 km $s^{-1}$ are in blue, and for speeds above 570 km $s^{-1}$ are in orange. In the second panel the slopes for OMNI in the range from 330 to 450 kms$^{-1}$ are in purple, and the slopes for speeds less than 450 km $s^{-1}$ are in green. Similarly for *ACE* the slopes in the range from 330 to 450 km $s^{-1}$ are in purple, and the slopes for speeds less than 450 km $s^{-1}$ are in green. The bottom panel shows the monthly sunspot number in black the average yearly number in pink, and the current sheet tilt is in blue.

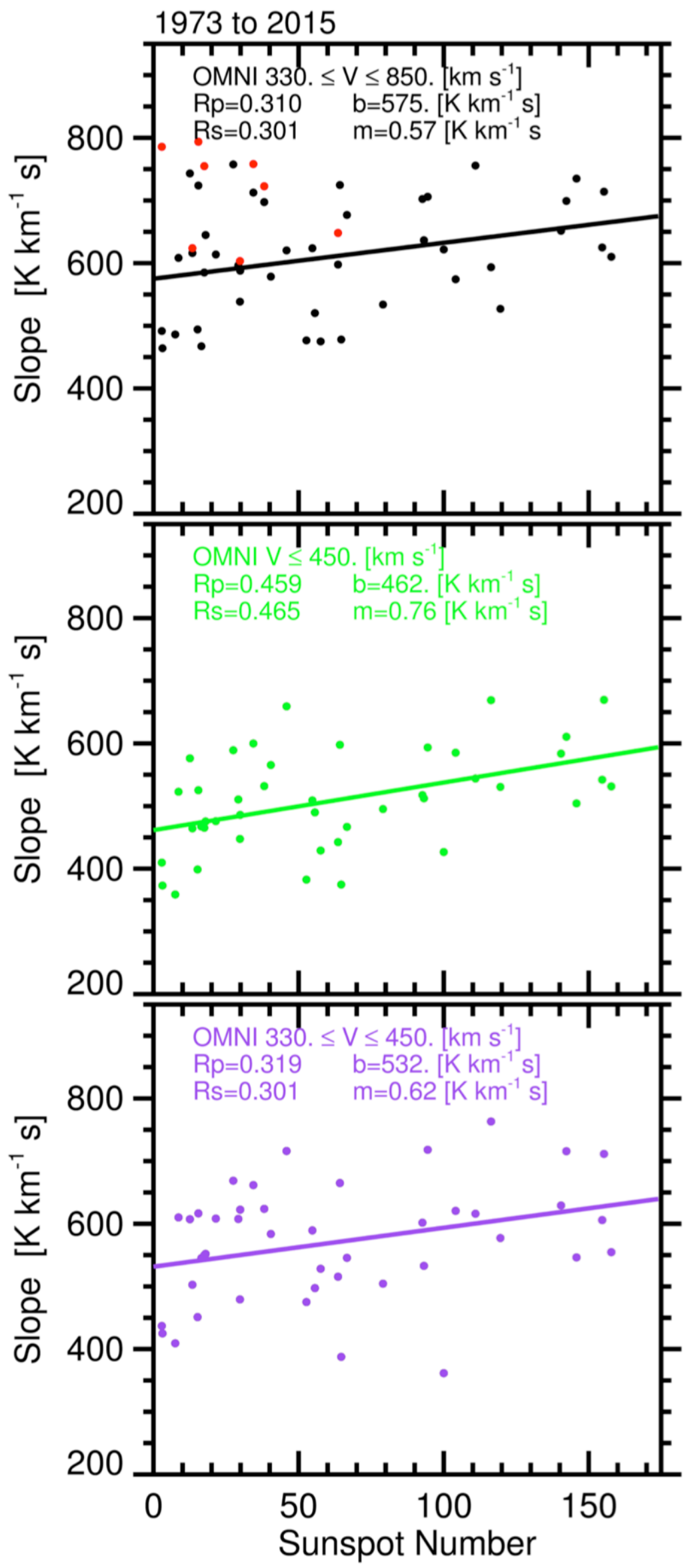

**Figure 3.** Scatter plots of OMNI yearly $T_p$-$V_p$ slopes versus the sunspot number where each panel corresponds to a different speed range: the top panel for speeds between 330 and 850 km $s^{-1}$, the middle panel for speeds ≤ 450 km $s^{-1}$, and the bottom panel for speeds ranging from 330 to 450 km $s^{-1}$. The solid lines are linear fits. The slopes (m) and intercepts (b) for the fits are labeled as well as the corresponding Pearson (Rp) and Spearman (Rs) correlation coefficients.

Table 1
Table of correlation coefficients, significance of the correlation coefficients, and slopes and intercepts for the analysis of the T-V slope and sunspot number data shown in Figure 3.

| | $330 \le V_p \le 850$ [km s$^{-1}$] | $V_p \le 450$ [km s$^{-1}$] | $330 \le V_p \le 450$ [km s$^{-1}$] |
|---|---|---|---|
| Spearman Correlation | 0.301 | 0.465 | 0.301 |
| Spearman Significance | 0.050 | $1.701\times10^{-3}$ | 0.050 |
| Pearson Correlation | 0.310 | 0.459 | 0.319 |
| Rcl | 1.017 | 1.506 | 1.045 |
| Slope [K km$^{-1}$ s] | 0.574 | 0.760 | 0.621 |
| Intercept [K km$^{-1}$ s] | 575.1 | 461.6 | 531.5 |

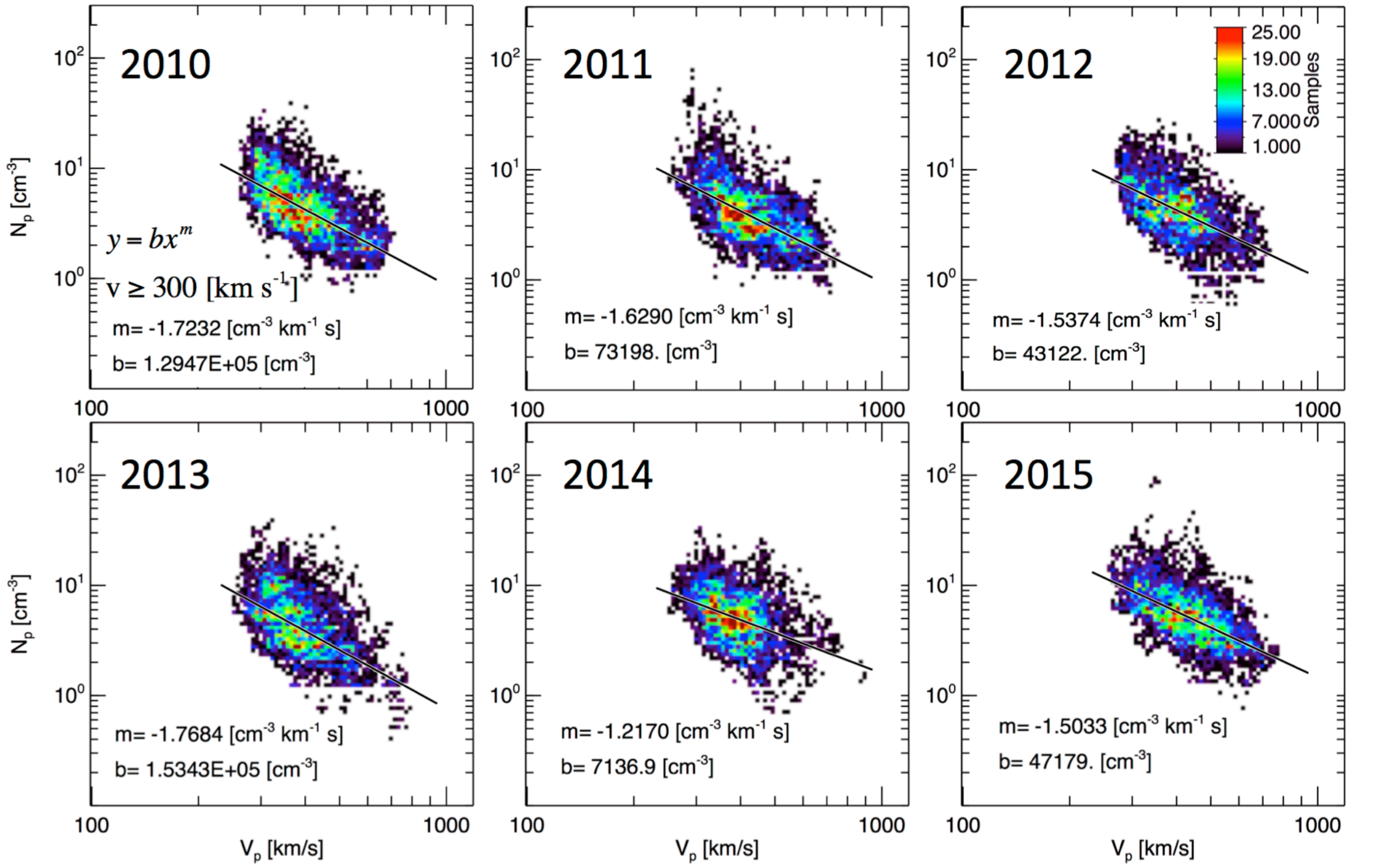

**Figure 4.** This is a collection of density-speed distribution plots spanning from 2010-2015. The analysis and format are similar to that in Figure 1. Key differences are that the fits are power law fits and both axes are logarithmic.

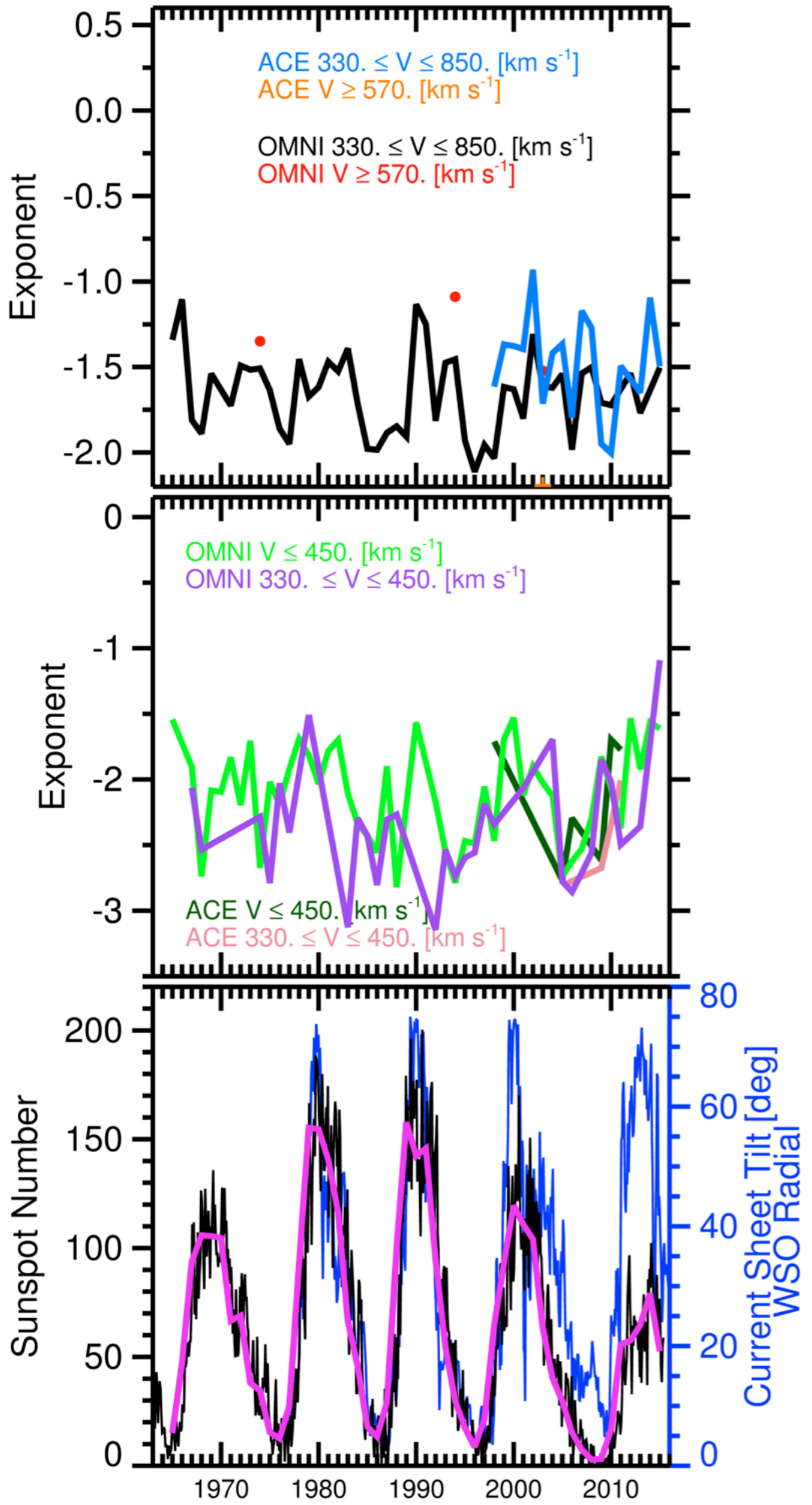

**Figure 5.** This plot has the same format as Figure 2, but instead of the $T_p$-$V_p$ slope the $n_p$-$V_p$ exponent for the power law fits are in the top 2 panels. The color-coding and speed ranges are the same as Figure 2.

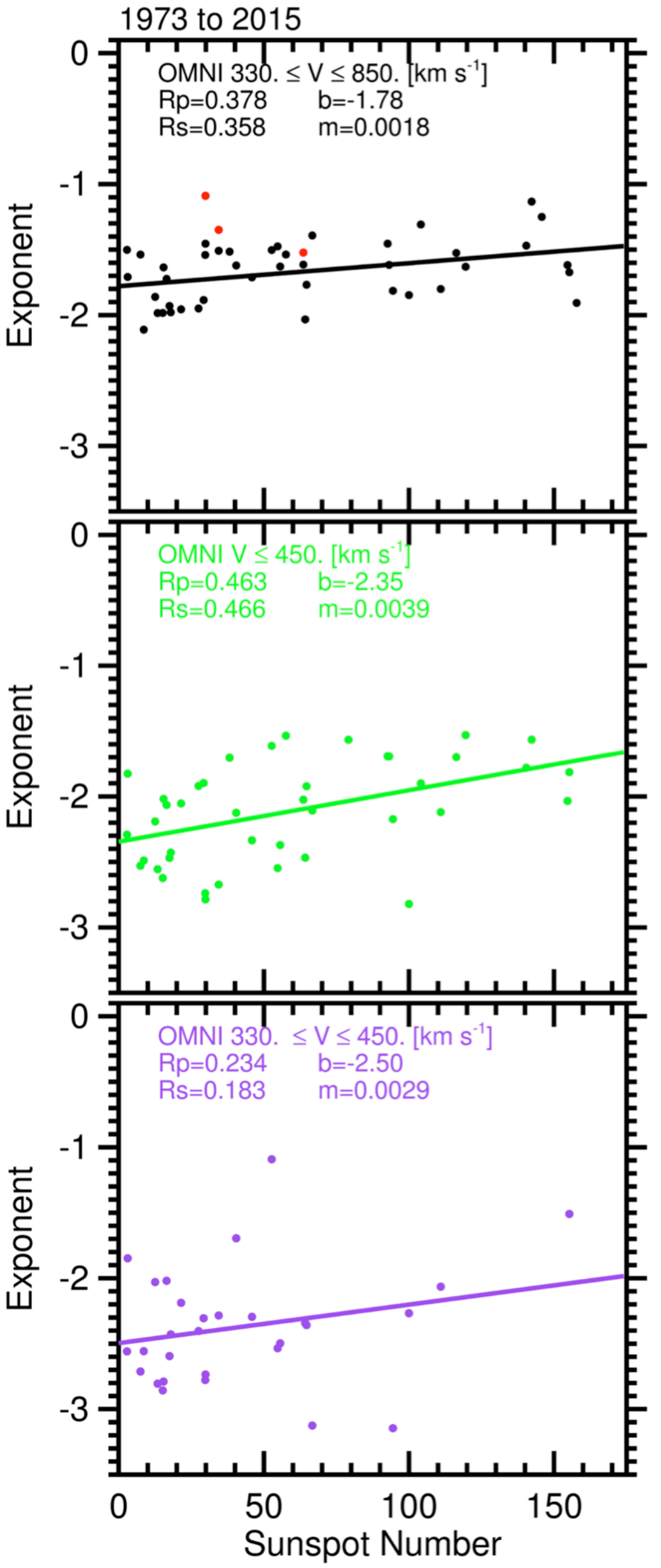

**Figure 6.** This figure has the same format as Figure 3, but instead of the $T_p$-$V_p$ slope the $n_p$-$V_p$ exponent is on the y-axis.

Table 2
Table of correlation coefficients, significance of the correlation coefficients, and slopes and intercepts for the analysis of the n-V exponent and sunspot number data shown in Figure 6.

| | $330 \leq V_p \leq 850$ [km s $^{-1}$] | $V_p \leq 450$ [km s $^{-1}$] | $330 \leq V_p \leq 450$ [km s $^{-1}$] |
|---|---|---|---|
| Spearman Correlation | 0.358 | 0.466 | 0.183 |
| Spearman Significance | 0.0198 | $2.125\times10^{-3}$ | 0.334 |
| Pearson Correlation | 0.378 | 0.463 | 0.235 |
| Rcl | 1.224 | 1.483 | 0.631 |
| Slope | $1.760\times10^{-3}$ | $3.931\times10^{-3}$ | $2.941\times10^{-3}$ |
| Intercept | -1.779 | -2.346 | -2.496 |

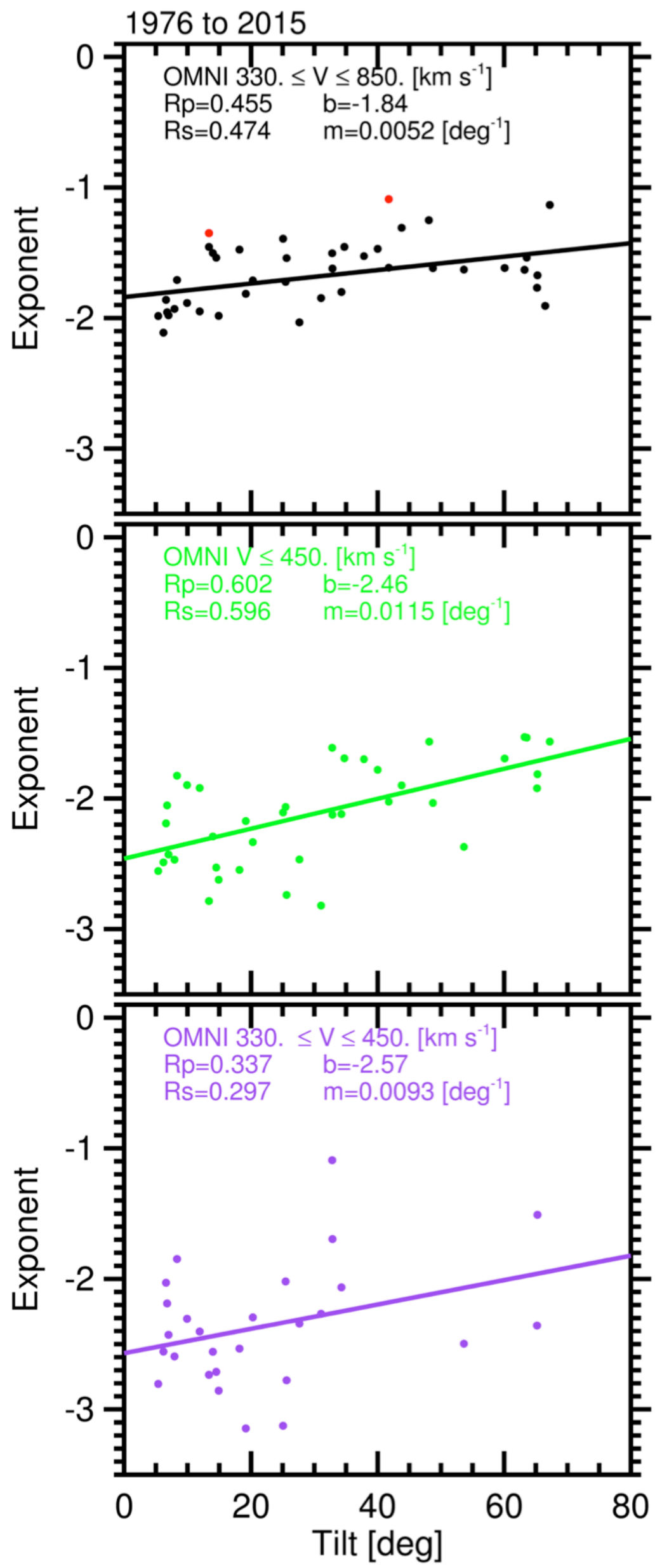

**Figure 7.** This figure has the same format as Figure 6, but the x-axis is the current sheet tilt angle instead of the sunspot number. This plot covers 1976 to 2015 since the tilt data began in 1976.

Table 3
Table of correlation coefficients, significance of the correlation coefficients, and slopes and intercepts for the analysis of the $n_p$-$V_p$ exponent and current sheet tilt angle data shown in Figure 7.

| | $330 \leq V_p \leq 850$ [km s$^{-1}$] | $V_p \leq 450$ [km s$^{-1}$] | $330 \leq V_p \leq 450$ [km s$^{-1}$] |
|---|---|---|---|
| Spearman Correlation | 0.474 | 0.596 | 0.297 |
| Spearman Significance | $2.320\times10^{-3}$ | $7.870\times10^{-5}$ | 0.1320 |
| Pearson Correlation | 0.455 | 0.602 | 0.337 |
| Rcl | 1.421 | 1.854 | 0.876 |
| Slope | $5.163\times10^{-3}$ | 0.0115 | 0.0094 |
| Intercept | -1.839 | -2.461 | -2.570 |

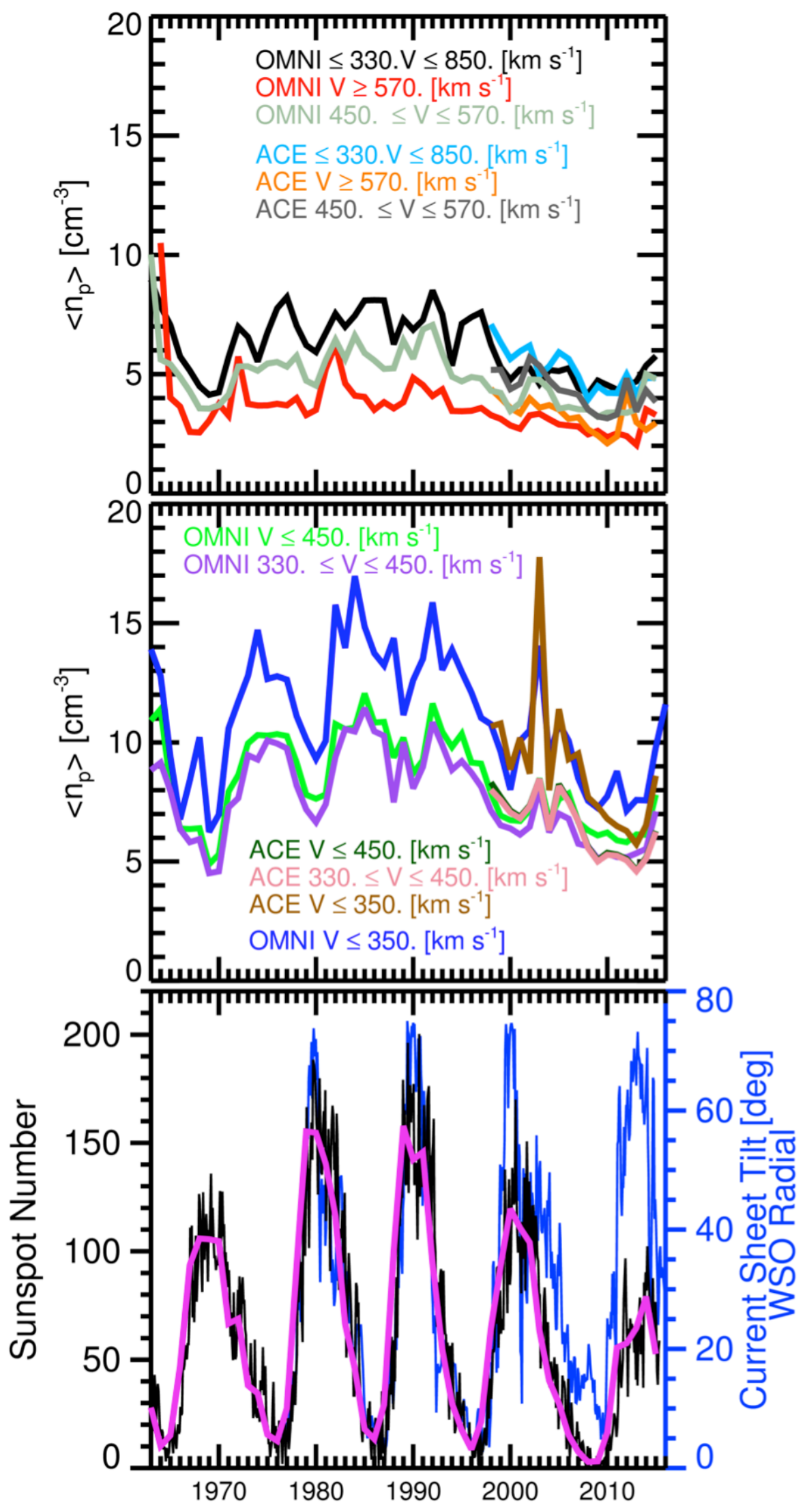

**Figure 8.** Similar format as Figure 1 with the yearly average proton density on the y-axis of the top and middle panels with additional results for data for speeds ≤ 350 km s$^{-1}$, and speeds between 450 and 570 km s$^{-1}$.

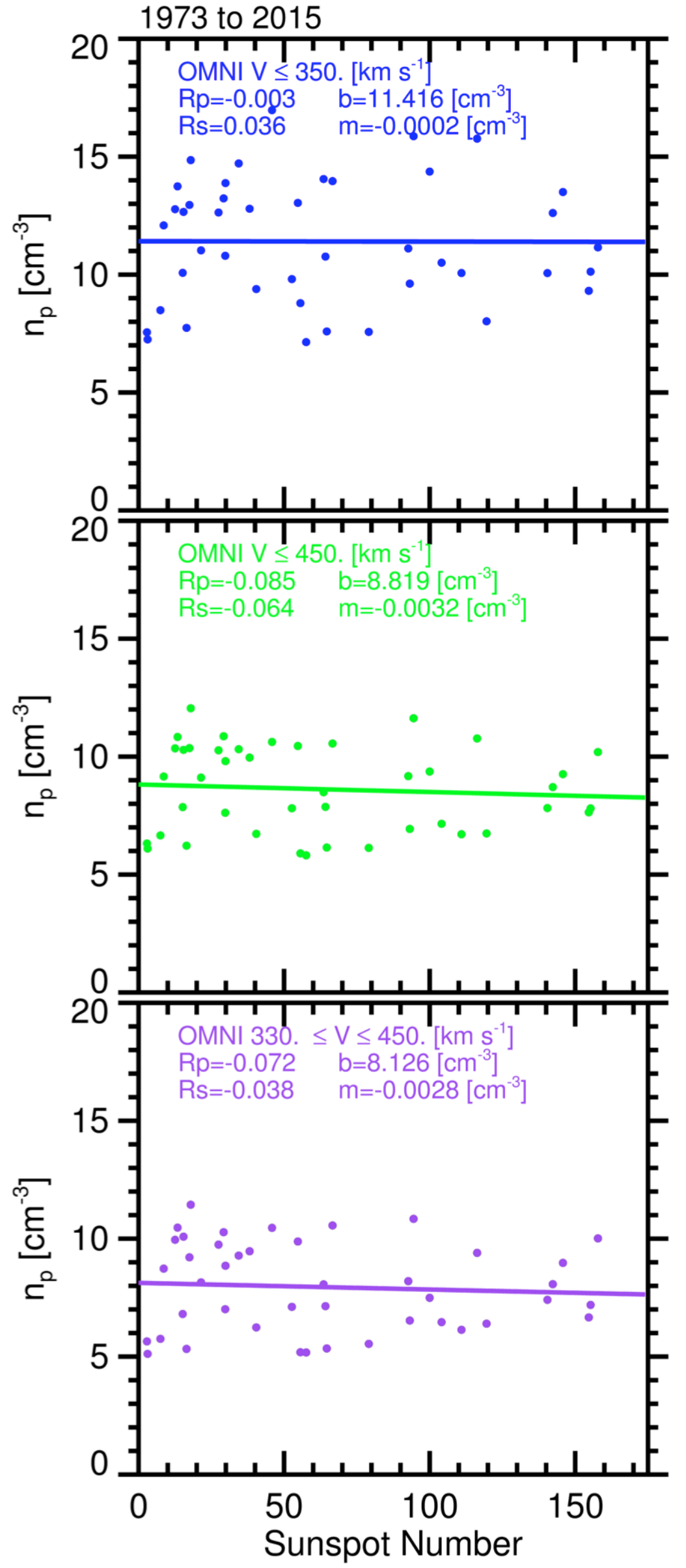

**Figure 9.** Similar format as Figure 3, but instead of the $T_p$-$V_p$ slope on the y-axis the the average yearly proton density is on the y-axis and the top panel shows results for speeds ≤ 350 km s$^{-1}$.

Table 4
Table of correlation coefficients, significance of the correlation coefficients, and slopes and intercepts for the proton density and sunspot data shown in Figure 9.

| | $V_p \leq 350$ [km s $^{-1}$] | $V_p \leq 450$ [km s $^{-1}$] | $330 \leq V_p \leq 450$ [km s $^{-1}$] |
|---|---|---|---|
| Spearman Correlation | 0.036 | -0.064 | -0.038 |
| Spearman Significance | 0.818 | 0.683 | 0.808 |
| Pearson Correlation | $2.971\times10^{-3}$ | -0.0849 | -0.0721 |
| Rcl | -0.010 | -0.278 | -0.237 |
| Slope [cm $^{-3}$] | $-1.625\times10^{-4}$ | $-3.195\times10^{-3}$ | $-2.810\times10^{-3}$ |
| Intercept [cm $^{-3}$] | 11.416 | 8.8190 | 8.1257 |

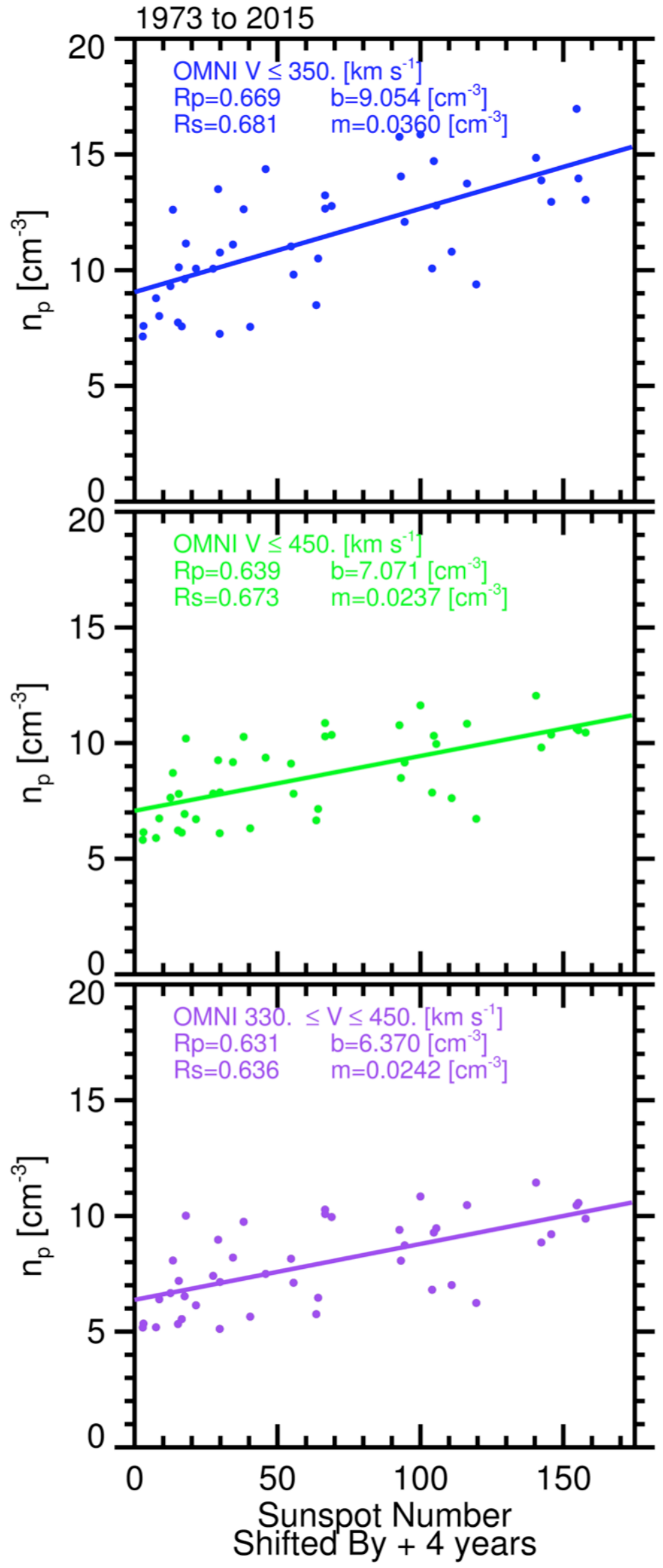

**Figure 10.** Same format and calculations as Figure 9 except that the sunspot data was time shifted by 4 years forward in time.

Table 5
Table of correlation coefficients, significance of the correlation coefficients, and slopes and intercepts for the proton density and time shifted sunspot data shown in Figure 10.

| | $V_p \leq 350$ [km s$^{-1}$] | $V_p \leq 450$ [km s$^{-1}$] | $330 \leq V_p \leq 450$ [km s$^{-1}$] |
|---|---|---|---|
| Spearman Correlation | 0.681 | 0.673 | 0.636 |
| Spearman Significance | $4.96\times10^{-7}$ | $7.50\times10^{-7}$ | $4.66\times10^{-6}$ |
| Pearson Correlation | 0.669 | 0.639 | 0.631 |
| Rcl | 2.194 | 2.095 | 2.067 |
| Slope [cm$^{-3}$] | 0.0360 | 0.0237 | 0.0242 |
| Intercept [cm$^{-3}$] | 9.054 | 7.071 | 6.370 |

Table 6
Table of correlation coefficients, significance of the correlation coefficients, and slopes and intercepts for the proton number flux and time shifted sunspot data shown.

| | $V_p \leq 350$ [km s$^{-1}$] | $V_p \leq 450$ [km s$^{-1}$] | $330 \leq V_p \leq 450$ [km s$^{-1}$] |
|---|---|---|---|
| Spearman Correlation | 0.691887 | 0.660 | 0.632 |
| Spearman Significance | 2.771E-07 | 1.444E-06 | 5.410E-06 |
| Pearson Correlation | 0.683 | 0.64194299 | 0.628 |
| Rcl | 2.240 | 2.1047508 | 2.059 |
| Slope [s$^{-1}$cm$^{-2}$] | 1.233E-06 | 9.2624E-05 | 9.293E-05 |
| Intercept [s$^{-1}$cm$^{-2}$] | 2.873E+08 | 2.552E+08 | 2.439E+08 |

Table 7
Table of correlation coefficients, significance of the correlation coefficients, and slopes and intercepts for the proton dynamic pressure and time shifted sunspot data shown.

| | $V_p \leq$ 350 [km s $^{-1}$] | $V_p \leq$ 450 [km s $^{-1}$] | 330 $\leq V_p \leq$ 450 [km s $^{-1}$] |
|---|---|---|---|
| Spearman Correlation | 0.702 | 0.642 | 0.624 |
| Spearman Significance | 1.540E-07 | 3.441E-06 | 7.743E-06 |
| Pearson Correlation | 0.690 | 0.634 | 0.622 |
| Rcl | 2.262 | 2.078 | 2.039 |
| Slope [nPa] | 7.009E-03 | 6.053 E-03 | 6.003E-03 |
| Intercept [nPa] | 1.530 | 1.564 | 1.571 |